\documentclass[%
 aip,
 amsmath,amssymb,
 reprint,%
]{revtex4-1}
 
\usepackage{graphicx}
\usepackage{dcolumn}
\usepackage{bm}

\usepackage[utf8]{inputenc}
\usepackage[T1]{fontenc}
\usepackage{mathptmx}
\usepackage{etoolbox}

\usepackage[utf8]{inputenc}
\usepackage[T1]{fontenc}

\usepackage[version=3]{mhchem} 
\usepackage{amsmath}
\usepackage{amsthm}
\usepackage{amsfonts}
\usepackage{amssymb}
\usepackage{mathtools}
\usepackage{chemformula}
\usepackage{multirow}
\usepackage{pythonhighlight}
\usepackage{makecell}

\theoremstyle{plain}
\newtheorem{theorem}{Theorem}

\newtheorem{lemma}[theorem]{Lemma}

\theoremstyle{definition}
\newtheorem{definition}[theorem]{Definition}

\theoremstyle{remark}

\theoremstyle{example}

\usepackage{titlesec}

\begin{document}

\newcommand{\nyu}{\affiliation{Simons Center for Computational Physical Chemistry, Department of Chemistry, New York University, New York, N.Y. 10003}}

\title{On the design space between molecular mechanics and machine learning force fields}

\author{Yuanqing~Wang}
\email{wangyq@wangyq.net}
\nyu
\affiliation{Center for Data Science, New York University, New York, N.Y. 10004}
\affiliation{Courant Institute of Mathematical Sciences, New York University, New York, N.Y. 10003}

\author{Kenichiro~Takaba}
\affiliation{Asahi Kasei Pharma Corporation, 632-1 Mifuku, Izunokuni, Shizuoka 410-2321, Japan
}

\author{Michael S. Chen}
\nyu

\author{Marcus~Wieder}
\affiliation{Open Molecular Software Foundation, Davis, Calif. 95616}

\author{Yuzhi~Xu}
\nyu
\affiliation{
NYU-ECNU Center for Computational Chemistry and \\
\mbox{Shanghai Frontiers Science Center of Artificial Intelligence and Deep Learning, NYU Shanghai, Shanghai 200062, P.R.China}}

\author{Tong~Zhu}
\affiliation{
NYU-ECNU Center for Computational Chemistry and \\
\mbox{Shanghai Frontiers Science Center of Artificial Intelligence and Deep Learning, NYU Shanghai, Shanghai 200062, P.R.China}}

\author{John~Z.~H.~Zhang}
\affiliation{
NYU-ECNU Center for Computational Chemistry and \\
\mbox{Shanghai Frontiers Science Center of Artificial Intelligence and Deep Learning, NYU Shanghai, Shanghai 200062, P.R.China}}
\affiliation{\mbox{Faculty of Synthetic Biology, Shenzhen University of Advanced Technology, Shenzhen 518055, P.R.China}}

\author{Arnav~Nagle}
\affiliation{University of Washington, Seattle, Wash. 98195}

\author{Kuang~Yu}
\affiliation{\mbox{Institute of Materials Research, Tsinghua Shenzhen International Graduate School, Tsinghua Univ., Shenzhen 518055, P.R.China}}

\author{Xinyan Wang}
\affiliation{DP Technology, Beijing, 100089, P.R.China}

\author{Daniel~J.~Cole}
\affiliation{\mbox{School of Natural and Environmental Sciences,
Newcastle University, Newcastle upon Tyne, NE1 7RU, U.K.}}

\author{Joshua~A.~Rackers}
\affiliation{Prescient Design, Genentech, New York, N.Y. 10004}

\author{Kyunghyun~Cho}
\affiliation{Center for Data Science, New York University, New York, N.Y. 10004}
\affiliation{Courant Institute of Mathematical Sciences, New York University, New York, N.Y. 10003}
\affiliation{Prescient Design, Genentech, New York, N.Y. 10004}


\author{Joe~G.~Greener}
\affiliation{\mbox{Medical Research Council Laboratory of Molecular Biology, Cambridge, CB2 0QH, U.K.}}

\author{Peter~Eastman}
\affiliation{Department of Chemistry, Stanford University, Stanford, Calif. 94305}

\author{Stefano~Martiniani}
\nyu
\affiliation{\mbox{Center for Soft Matter Research, Department of Physics, New York University, New York, N.Y. 10003}}
\affiliation{Courant Institute of Mathematical Sciences, New York University, New York, N.Y. 10003}

\author{Mark~E.~Tuckerman}
\nyu
\affiliation{Courant Institute of Mathematical Sciences, New York University, New York, N.Y. 10003}

%
%
\begin{abstract}
A force field as accurate as quantum mechanics (QM) and as fast as molecular mechanics (MM), with which one can simulate a biomolecular system efficiently enough and meaningfully enough to get quantitative insights, is among the most ardent dreams of biophysicists---a dream, nevertheless, not to be fulfilled any time soon.
Machine learning force fields (MLFFs) represent a meaningful endeavor towards this direction, where differentiable neural functions are parametrized to fit \textit{ab initio} energies, and furthermore forces through automatic differentiation.
We argue that, as of now, the utility of the MLFF models is no longer bottlenecked by accuracy but primarily by their speed (as well as stability and generalizability), as many recent variants, on limited chemical spaces, have long surpassed the \textit{chemical accuracy} of $1$ kcal/mol---the empirical threshold beyond which realistic chemical predictions are possible---though still magnitudes slower than MM.
Hoping to kindle explorations and designs of faster, albeit perhaps slightly less accurate MLFFs, in this review, we focus our attention on the design space (the speed-accuracy tradeoff) between MM and ML force fields.
After a brief review of the building blocks of force fields of either kind, we discuss the desired properties and challenges now faced by the force field development community, survey the efforts to make MM force fields more accurate and ML force fields faster, and envision what the next generation of MLFF might look like.
\end{abstract}

\maketitle

%
%
\section{Introduction: The past, present, and future of force fields.}

\begin{figure*}
    \centering
    \includegraphics[width=0.6\linewidth]{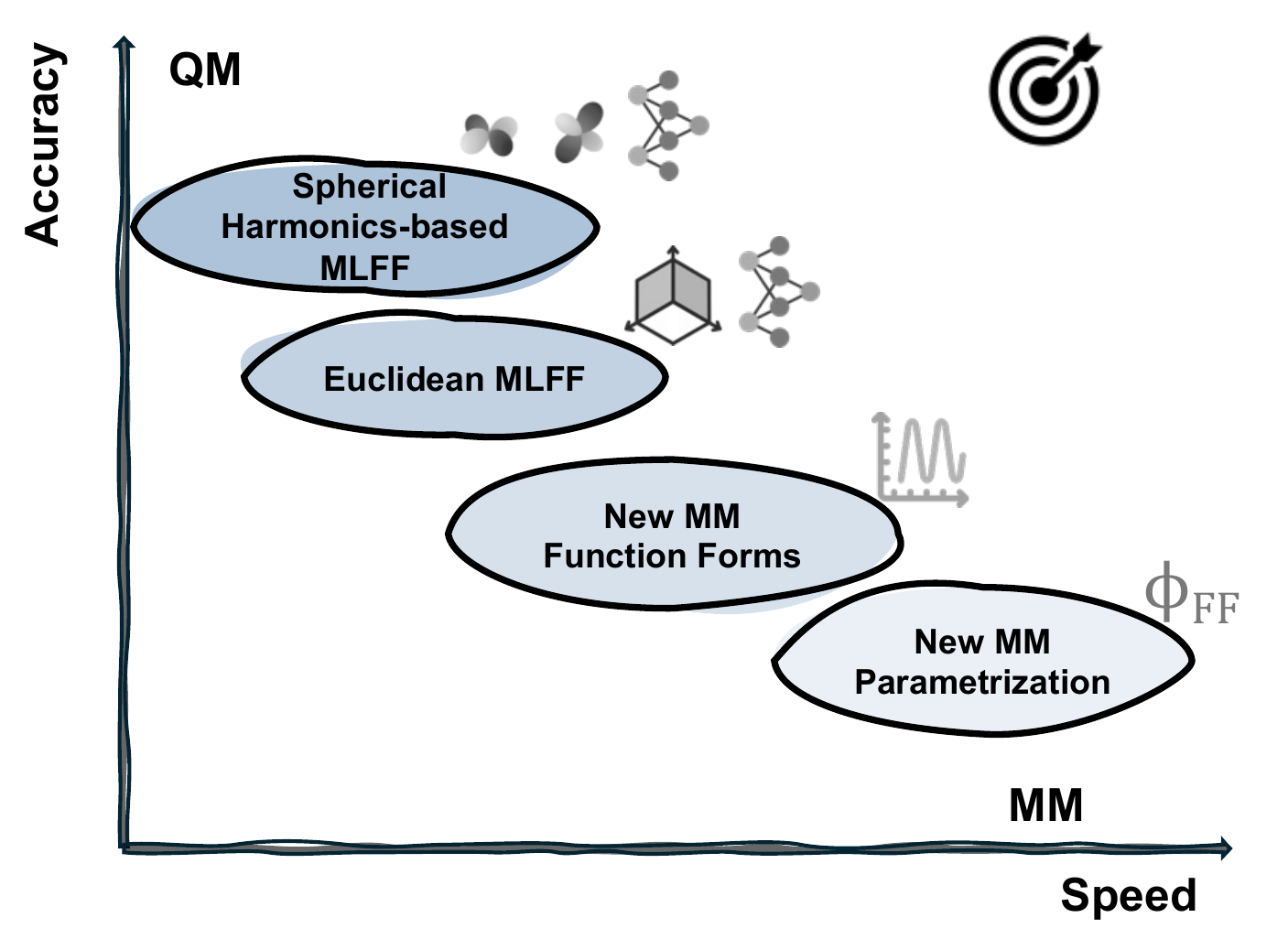}
    \caption{\textbf{Overview of the design space} between molecular mechanics (MM) and machine learning (ML) force fields.}
    \label{fig:abstract}
\end{figure*}

Computational chemists and biophysicists are interested in characterizing the \textit{energy landscape} of many-body systems---the distribution of coordinates $\mathbf{x} \in \mathbb{R}^{N \times 3}$ at a certain state with energy $U(\mathbf{x})$, which adopts a Boltzmann form~\cite{boltzmann1868studien}:
\begin{equation}
\label{eq:boltzmann}
p(\mathbf{x}) \propto \exp \left( -\frac{U(\mathbf{x})}{kT} \right),
\end{equation}
where $N$ is the number of particles in a many-body system, $k$ the Boltzmann constant, and $T$ the temperature.
In a simulation, since the true (reference) energy $U$ is almost always inaccessible, and \textit{ab initio} methods are usually prohibitively expensive, one resorts to a surrogate model which models the energy landscape as a function based upon the coordinates $\mathbf{x}$, the identity of the particles in the system $\mathbf{h} \in \mathbb{R}^{N}$, and a set of parameters $\Phi$:
\begin{equation}
\label{eq:forcefield}
\hat{p} \propto \exp \left( -\frac{\hat{U}(\mathbf{x}; h, \Phi)}{kT} \right).
\end{equation}
Evidently, the closer $U$ and $\hat{U}$ are, the smaller the divergence between the true and simulated probability distribution $p$ and $\hat{p}$ will be.
We call this parametrized scalar field $\hat{U}: \mathbb{R}^{N \times 3} \rightarrow \mathbb{R}^{1}$ a \textit{force field} (FF).
Molecular dynamics (MD) is usually employed to generate samples from this distribution.

\begin{figure}
    \centering
    \includegraphics[width=\linewidth]{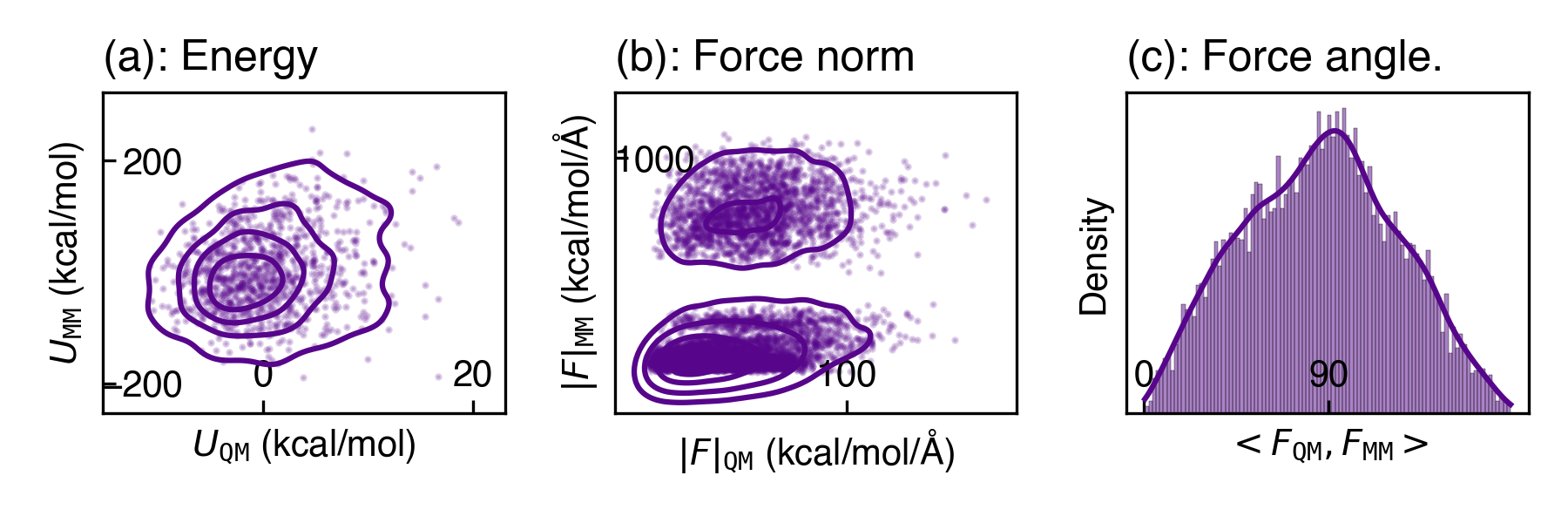}
    \caption{
    \textbf{Between MM and QM energies and forces, there is little correlation.}
    Scatter plots and kernel density estimate (KDE) of:
    (a): MM energy ($U_\mathtt{MM}$, mean-subtracted) plotted against QM energy ($U_\mathtt{QM}$, mean-subtracted);
    (b): MM force magnitude ($|F|_\mathtt{MM}$) plotted against QM force magnitude ($|F|_\mathtt{QM}$);
    (c): Distribution of deviation of angles between QM and MM forces.
    QM energies refer to the CCSD(T) computation of the ethanol molecule in MD17~\cite{chmiela2017} dataset.
    MM energies and forces are re-calculated using the state-of-the-art \texttt{openff-2.0.0}~\cite{boothroyd2023development} force field. 
    }
    \label{fig:qm-mm}
\end{figure}

Dating back to \citet{mccammon1977dynamics} in \citeyear{mccammon1977dynamics}, molecular mechanics (MM) force fields have been curated, using structural and QM data, to capture the qualitative behavior of biomolecular systems~\cite{ponder2003force, van2005gromacs, case2005amber, phillips2005scalable, calculations2007theory, WANG2014979, li2015molecular, sun2016compass} to power the \textit{in silico} modeling of all aspects of chemistry, from drug discovery to material sciences.
The blessing and the curse of MM force fields both lie in their simple functional forms (Equation~\ref{eq:u_mm}).
On the one hand, these terms afford linear $\mathcal{O}(N)$ runtime complexity and can be further aggressively optimized in modern compute hardware namely graphics processing units (GPU), simulating more than hundreds of nanoseconds per day for many biomolecular drug targets~\cite{harvey2009acemd,salomon2013routine,eastman2017openmm, eastman2023openmm} while still achieving useful accuracy for tasks such as predicting protein-ligand binding free energies~\cite{wang2015accurate,schindler2020large,gapsys2022pre}.
At the same time, the limited \textit{expressiveness} of this functional form dictates that it is impossible to fit the QM energies and forces well, especially in the high-energy region---see Figure~\ref{fig:qm-mm} for a comparison of QM and MM energies of a very simple ethanol molecule in the MD17~\cite{chmiela2017machine} dataset, although they recover the \textit{position} of QM minima relatively well~\cite{boothroyd2023development}.
Worse still, even within the limits of this functional form, there is no guarantee that maximal expressiveness has been achieved, as the assignment of parameters to the chemical motifs (atoms, bonds, angles, and torsions) in a MM force field has been relying on a human-derived, labor-intensive, and inextendable scheme termed \textit{atom typing}, where atoms of distinct nature are forced to share parameters.
\citet{takaba2023machinelearned} shows that, on very limited chemical space and low energy region, the energy disagreement between legacy force fields and QM is far beyond the chemical accuracy of 1 kcal/mol---the empirical threshold under which we believe that the qualitative characterization of a many-body system is possible.
And even when coupled with a trainable, flexible parametrization engine, the \textit{training} accuracy still cannot exceed the chemical accuracy.
Limitations exist in both the \textit{functional form} and \textit{parametrization} steps of MM force fields.

Another line of fruitful research~\cite{unke2021machine, smith2017ani,smith2018less,smith2019approaching,devereux2020extending,schutt2018schnet,batzner2021se,10.1093/bib/bbab158, wang2023spatial, musaelian2022learning, defabritiis2024machinelearningpotentialsroadmap, barnett2024neural} focuses on developing machine learning force fields (MLFFs) with flexible functional forms to fit the energy (and force) landscape of \textit{ab initio} calculations.
Typically, the energy is predicted by a (typically equivariant or invariant) neural network, whereas the forces are given by using automatic differentiation w.r.t. the positions of particles.
We argue that the energy and force accuracy of machine learning force fields is no longer a limiting factor for the wide applications thereof, as most such models achieve an energy error well below the chemical accuracy of $1$ kcal/mol.
Besides stability and generalizability, the speed of MLFFs, is what prevents them from wide applications---although they are usually by magnitudes faster than QM calculations (and scale linearly w.r.t. the size of the system), they are still hundreds of times slower than MM force fields.
For small molecule systems up to $10^2$ atoms, some of the fastest MLFFs~\cite{wang2023spatial} still take around 1 millisecond per energy and force evaluation on an A100 GPU, compared to less than 0.005 milliseconds for MM force fields.
As such, to simulate any biomolecular system of considerable size for a reasonably long time frame would usually have a prohibitive computational cost, significantly limiting its deployment to further our understanding of biologically relevant systems.

In this review, we direct our attention to the design space (in terms of the speed-accuracy tradeoff) between ML and MM force fields.
Hoping that this would inspire the design of a class of MLFF incorporating MM philosophy that is significantly faster, more stable, more interpretable, and more generalizable, than current state-of-the-art MLFFs albeit slightly less accurate/expressive, we organize this review as follows:
First, in Section~\ref{sec:desiderata}, we outline the \textit{desiderata} of force fields.
Secondly, in Section~\ref{sec:mm} and Section~\ref{sec:toolbox} we briefly review the building blocks of both MM and ML force fields and recent advances in the functional forms and parametrizations thereof.
Finally, after a discussion surrounding the datasets and best practices for fitting MM and ML force fields in Section~\ref{sec:data}, we envision the shape of the next generation of ultra-fast MLFF with high utility in biomolecular modeling in Section~\ref{sec:dream}.
We summarize (albeit with considerable overgeneralization) the key theoretical and practical properties discussed in this paper in Table~\ref{tab:comparison}.

\begin{table*}[htbp]
    \centering
    \begin{tabular}{c c c}
    \hline
    & Molecular mechanics (MM) & Machine learning force fields (MLFF) \\
    \hline
    Genesis & \citet{mccammon1977dynamics}(\citeyear{mccammon1977dynamics})
    & \citet{PhysRevLett.98.146401}(\citeyear{PhysRevLett.98.146401})\\
    Runtime Complexity & $\mathcal{O}(N)$ & $\mathcal{O}(N)$\\
    Speed & $> 1 \mu $s / day & around $1$ ns / day\\
    Invariance & $E(3)$ & $E(3)$ \\
    Equivariant universality & Impossible & Possible \\
    Accuracy & $>1$ kcal/mol & $<< 1$ kcal/mol for small molecules\\
    Stability & Usually guaranteed & Not guaranteed \\
    Topology & Usually required & Usually not required \\
    Force differentiation & Analytical & Autograd \\
    Long-range interactions & Modeled & Usually ignored \\
    Parametrization & Human-derived & Automated \\
    Customization & Difficult & Trivial \\
    Platform & Specialized & Tensor-accelerating frameworks \\
    
    \hline
    \end{tabular}
    \caption{Molecular mechanics (MM) vs. machine learning force fields (MLFF).}
    \label{tab:comparison}
\end{table*}

\section{Desiderata: The balance between speed and accuracy, and beyond.}
\label{sec:desiderata}
The speed and accuracy are two natural axes the community is interested in (Figure~\ref{fig:abstract})---a faster and more accurate force field is almost always desired.
With this in mind, we review the desired properties of a force field, regardless of its type, for the applications in physical modeling.

\paragraph{Invariance. }
Symmetries are inherent in all physical systems.
To exploit these symmetries in model construction and parametrization enhances data efficiency as it avoids unnecessary replicated manifestations of the same piece of data and avoids unphysical model interpretations.
Formally, we use \textit{equivariance} and \textit{invariance} to describe the symmetry in functional space:
\begin{definition}[Equivariance and invariance]
A function $f: \mathcal{X} \rightarrow \mathcal{Y}$ is said to be \textit{equivariant} to a symmetry group $G$ if 
\begin{equation}
    f(T_g(\mathbf{x})) = S_g(f(\mathbf{x}))
\label{eq:eqvariance_definition}
\end{equation}
and \textit{invariant} if
\begin{equation}
    f(T_g(\mathbf{x})) = f(\mathbf{x})
\label{eq:invariance_definition}
\end{equation}
$\forall \,  g \in G$ and some equivalent transformations on the two spaces respectively $T_g: \mathcal{X} \rightarrow \mathcal{X}$ and $S_g: \mathcal{Y} \rightarrow \mathcal{Y}$.
\end{definition}
In the three-dimensional space we happen to dwell in, the true underlying energy function adopts $E(3)$-invariance, where the $T_g = S_g$ in Equation~\ref{eq:eqvariance_definition} and Equation~\ref{eq:invariance_definition} are the rotations ($T_g(\mathbf{x}) = \mathbf{x} R$, where $R \in \mathbb{R}^{3 \times 3}$ is a rotation matrix $R R^{T} = I$), translations ($T_g(\mathbf{x}) = \mathbf{x} + \Delta \mathbf{x}$), and reflections ($T_g(\mathbf{x}) = \operatorname{Ref}_\theta(\mathbf{x})$).
These hold since the coordinate system on which the Euclidean coordinates are based is artificial and arbitrary.
Similarly, since the indexing of the particles is arbitrary, permutation equivariance and invariance are also crucial: $T_g(\mathbf{x}) = S_g(\mathbf{x}) = P\mathbf{x}$, where $P$ is the associated permutation matrix.
We also point out that there have been recent efforts showing that non-equivariant/non-invariant models can also be trained to perform on equivariant tasks~\cite{thais2023equivariance, puny2022frameaveraginginvariantequivariant, duval2023faenetframeaveragingequivariant}.

\paragraph{Linear runtime complexity. }
One of the key applications of force fields in biomolecular simulations is modeling the energy landscapes of heterogeneous biomolecules, such as protein-ligand complexes.
In such cases, the number of atoms of such systems can significantly exceed $10^{3}$, making the training and inference unattainably expensive if the runtime complexity is anywhere higher than quadratic $\mathcal{O}(N^2)$.
At first glance, it might not be intuitive that the popular force fields, including MM (Equation~\ref{eq:u_mm}), which have pairwise interactions, are linear.
In practice, however, not all $N^2$ pairwise interactions are present in the energy function, as a \textit{cutoff} function is almost always employed to mask off any interaction with a distance longer than a threshold $L$:
\begin{equation}
\label{eq:cutoff}
\lambda(\mathbf{x}_1, \mathbf{x}_2) =
\begin{cases}
\Lambda(||\mathbf{x}_1 - \mathbf{x}_2||), 
||\mathbf{x}_1 - \mathbf{x}_2|| < L, \\
0, \mathtt{elsewhere},
\end{cases}
\end{equation}
with smooth boundary condition $\lim\limits_{||\mathbf{x}_1 - \mathbf{x}_2|| \rightarrow L} \Lambda = 0$.
For both ML force fields, $\Lambda$ usually is typically implemented in the form of a cosine annealing;
a reaction field method~\cite{barker1973monte, watts1974monte} is typically used for Coulomb interactions whereas a polynomial switching function is used for van der Waals interactions.
This goes hand-in-hand with the assumption of constant sparsity: on average, the number of interactions that fall within the cutoff boundary for each particle stays constant.
(Note that this is a strong assumption that would not stand without prior knowledge, i.e. $p(x)$ adopts a uniform distribution.)

\paragraph{Energy conservation. }
Since the potential energy $U(x)$ is a state function, there is zero work associated with moving a particle through a trajectory starting and ending at the same place~\cite{chmiela2017machine}: 
\begin{equation}
\label{eq:zero-work}
W \equiv \oint_C \vec{F} \cdot \mathrm{d}\vec r = 0.
\end{equation}
This is satisfied if the force is always implemented as the negative gradient of the energy:
\begin{equation}
F = -\nabla_x U(\mathbf{x}).
\end{equation}
Numerically, Equation~\ref{eq:zero-work} always stand accurately for simple functional forms such as an MM force field, but less so for ML force fields with sophisticated functional forms.

\paragraph{Differentiability. }
In the previous paragraph, we have discussed the differentiation of the energy $\hat{U}(\mathbf{x}; h, \Phi)$ w.r.t. $\mathbf{x}$, which yields the force prediction.
In MLFF, to ensure that at least the first- and second-order derivatives are smooth and further differentiable, neural activation function choices are limited to those with continuous first- and second-order derivatives, such as the $\operatorname{ELU}$~\cite{clevert2016fast}-family.
Meanwhile, if one wishes to optimize the MM force field parameters $\Phi_\mathtt{FF}$, it is also crucial to evaluate the parameter derivative $\partial \hat{U}(\mathbf{x}; h, \Phi_\mathtt{FF}) / \partial \Phi_\mathtt{FF}$.

\paragraph{Universality. }
Typically, there is little prior knowledge to be encoded in the inductive bias about the \textit{ab initio} energy landscape except for the fact that when particles are separated far enough, the energy approaches zero.
As such, at least in a limited region, the force field should be expressive enough to represent a diverse set of functions, or ideally \textit{universal}, modulo the group invariance.
\begin{definition}[Universality of invariant functions]
A parametrized invariant function $f(\cdot; \Phi)$ is said to be universal w.r.t. a symmetry group $G$ on a space $\mathcal{X}$ if for all transformations in that symmetry group $\forall T \in G$, and for all invariant functions $g$ satisfying $g(T(\mathbf{x})) = g(\mathbf{x})$, there exists a set of parameters $\Phi$ such that $f$ can be approximated arbitrarily well:
\begin{eqnarray}
|| g(\mathbf{x}) - f(\mathbf{x}, \Phi) || < \epsilon, \forall \mathbf{x} \in \mathcal{X},
\end{eqnarray}
for arbitrarily small $\epsilon > 0$.
\end{definition}

\paragraph{Stability.}
\citet{fu2023forces} have illustrated that for many MLFF models, the simulation would collapse into unphysical regions after a certain number of integration steps.
From a data-centric perspective, this largely results from not having substantial high-energy samples in training.
Tackling this problem with sampling~\cite{stocker2022robust} on the high-energy regions, possibly with active learning~\cite{smith2018less, wang2020active, schwalbe2021differentiable}, is a data-driven way to counteract this instability, although this might lead to significantly more data as the high-energy regions (in a Tolstoyan way) are much more diverse.
On the other hand, MM force fields are almost intrinsically stable since the functional forms are designed to be restrictive.
The balance between expressiveness and stability resembles that between bias and variance, and given the same amount of data, can only be achieved with carefully designed inductive biases. 

\paragraph{How fast is fast enough?}
To be able to simulate one millisecond per day~\cite{eastman2017openmm} of macromolecular behavior is a target that modern GPU-accelerated MM frameworks strive to achieve.
As such, mechanisms such as protein folding, whose time scale is milliseconds~\cite{lindorff2011fast, kubelka2004protein, eaton2021modern, manavalan2019pfdb, horng2003rapid}, can be simulated within a reasonable time frame.
For small molecules, current ML force fields are about $10^2$ to $10^3$ slower than MM~\cite{wang2023spatial, qiao2020orbnet}.
Since they both scale linearly, it is not unreasonable to estimate that this ratio is similar for large systems.

When it comes to using these force fields to simulate biomolecular systems in a drug discovery setting, it is also worth paying attention to the economics of simulations~\cite{retchin2024druggym}.
Since the cost of one GPU hour is around \$1, whereas the cheapest wet lab assay, in a high-throughput screening setting only might cost less than \$100, it is highly likely that to reliably estimate an observable \textit{in silico} using ML force fields might cost significantly more than measuring it in a wet lab, provided that the acquisition of the molecule and the biological system is not extraordinarily expensive.

Finally, it is worth noting that most compute resources is spent on \textit{force}, rather than \textit{energy} evaluations for both MM and ML force fields, and \textit{energy} evaluations can even be skipped for intermediate steps.
With MM, however, the forces are computed with analytical gradients rather than automatic differentiation~\cite{baydin2018automatic}.
See more discussion in Section~\ref{subsec:ecosystem}.

\paragraph{How accurate is accurate enough?}
The \textit{chemical accuracy} of 1~kcal/mol has long been an empirical standard beyond which the estimation of key chemical properties consistent with experiments is qualitatively possible.
In a Nobel lecture~\cite{pople1999nobel}, \citeauthor{pople1999nobel} set this threshold as the target accuracy for the estimation of formation and ionization potentials.
We also note that the discrepancies among various levels of theory are higher than this threshold, especially in high-energy regions.
Moreover, the error of density functional theory (DFT) calculations also routinely exceed this number~\cite{bogojeski2020quantum}.

For biomolecular applications, if one were to apply a force field to calculate the binding free energies of a protein-ligand system, \citet{mobley2012perspective} have demonstrated that a 1~kcal/mol noise in the \textit{free energy} would roughly translate to making a drug discovery campaign 5 times faster if the goal is to increase the association constant $K_a$ by ten folds (admittedly, the noise in the free energy is not the same as the noise in the potential energy, but we provide this finding here as an empirical gauge).

In some limited chemical spaces, the state-of-the-art ML potentials have long surpassed this threshold.
Take the benchmark study~\cite{batatia2023mace, Batzner_2022} on (r)MD17~\cite{chmiela2017, christensen2020rolegradientsmachinelearning}, for example, almost all competitive models have error well below $10^{-1}$ kcal/mol on all molecules, and the differences among the top 3 models usually come down to the scale of $10^{-2}$ kcal/mol.
On a more diverse dataset containing organic small molecules~\cite{eastman2023spice, smith2020ani}, to achieve accuracy significantly higher than the quantum chemical threshold of 1 kcal/mol has long been possible~\cite{kovács2023maceoff23transferablemachinelearning, simeon2024tensornet}.

%
%
\section{Molecular mechanics force fields: Simple, crude, but practical for wide applications}

\label{sec:mm}
\subsection{Classical MM force fields}
Molecular mechanics (MM) force fields~\cite{dauber2019biomolecular, hagler2019force} are empirical models representing atomic point masses interacting through parametrized functions of atomic coordinates. These functions characterize the potential energy of a system via valence (bond, angle, and torsion) and nonbonded terms, typically expressed as the sum of polynomials and truncated Fourier series. 
The most popular and widely used MM force field in biomolecular modeling and simulation is the Class I MM force field, primarily due to its computational efficiency, arising from its simplified functional form which can be typically expressed as:
\begin{eqnarray}
\label{eq:u_mm}
    & U_\text{MM}(\mathbf{x}; \Phi_\mathtt{FF}) \\
    &= \sum\limits_{\text{bond}} & \frac{K_r}{2} (r_{ij} - r_0)^2 \nonumber \\
    &+ \sum\limits_{\text{angle}} & \frac{K_\theta}{2} (\theta_{ijk} - \theta_0)^2 \nonumber \\
    &+ \sum\limits_{\text{torsion}} & \sum\limits_{n=1}^{n_\text{max}} K_{\phi, n}\left[1 + \cos(n \phi_{ijkl} - \phi_0)\right] \nonumber \\
    &+ \sum\limits_{\text{Coulomb}} & \frac{1}{4 \pi \epsilon_0} \frac{q_i \, q_j}{r_{ij}} \nonumber \\
    &+ \sum\limits_{\text{van der Waals}} & 4 \epsilon_{ij} \left[ \left (\frac{\sigma_{ij}}{r_{ij}} \right)^{12} - \left(\frac{\sigma_{ij}}{r_{ij}}\right)^{6} \right] \nonumber,
\end{eqnarray}
where the total potential energy $U_\mathtt{MM}$ as a function of the coordinates of the system $\mathbf{x}$ and the collection of force field parameters (Also see Section~\ref{subsec:parametrization}) $\Phi_\mathtt{FF} = \{ K_r, K_\theta, r_0, \theta_0, K_{\phi, n}, \phi_0, q, \sigma, \epsilon\}_i$ is modeled as the sum of bond, angle, torsion, and nonbonded energy.
The bond, angle, and torsion force constants are represented as $K_r, K_\theta$ and $K_{\phi, n}$, with their equilibrium values and phases denoted as $r_0, \theta_0$ and $\phi_0$, respectively. The atomic point charges are represented by $q$, while $\epsilon$ and radii $\sigma$ parametrize the Lennard-Jones energy well.
$r_{i, j}, \theta_{i, j, k}, \phi_{i, j, k, l}$---which represents the distance between covalently bonded atoms $i, j$, the angle among $i-j-k$, and the dihedral angle between the planes formed by $i, j, k$ and $j, k, l$, respectively---are simple functions extracted from the coordinates $\mathbf{x}_{i, j, k, l}$.
In practice, such operations can be implemented as:
\begin{gather}
\label{eq:geometry}
r_{i, j} = || \vec{\mathbf{x}}_{ij}|| =  || \mathbf{x}_i - \mathbf{x}_j ||;\\
\theta_{i, j, k} = \operatorname{atan2}(
||\vec{\mathbf{x}}_{ij} \times \vec{\mathbf{x}}_{jk} ||, \vec{\mathbf{x}}_{ij} \cdot \vec{\mathbf{x}}_{jk}
);\nonumber\\
\phi_{i, j, k, l} = \operatorname{atan2}( \nonumber \\
((\vec{\mathbf{x}}_{ij} \times \vec{\mathbf{x}}_{jk}) \times (\vec{\mathbf{x}}_{jk} \times \vec{\mathbf{x}}_{kl})) \cdot \frac{\vec{\mathbf{x}}_{jk}}{|| \vec{\mathbf{x}}_{jk} ||},\nonumber\\
((\vec{\mathbf{x}}_{ij} \times \vec{\mathbf{x}}_{jk}) \cdot (\vec{\mathbf{x}}_{jk} \times \vec{\mathbf{x}}_{kl}))).\nonumber
\end{gather}

An out-of-plane term, known as the \textit{improper torsion}, can be also introduced with the torsion term to enhance the molecular planarity and prevent chiral inversions. 
In theory, multipole expansion---such as dipole and quadrupole moments---are necessary to accurately represent the quantum mechanical electrostatic potential. However, empirical force fields try to approximate this multipole expansion by assigning point charges localized at the nuclei of atoms, sometimes with virtual sites to model lone pairs and $\sigma$-holes~\cite{ringrose2022exploration, yan2017improved, kolar2016computer}, in order to reproduce the same electrostatic potential that would be given by the true electronic structure and electron density distribution. 
The van der Waals interaction combines repulsive and attractive forces, typically represented by a 12-6 Lennard-Jones potential. The Lorentz–Berthelot~\cite{delhommelle2001inadequacy} combining rules can be employed to determine $\sigma$ and $\epsilon$ between different atom types, though alternative combination rules are possible~\cite{halgren1992representation}.
The $r^{-12}$ term accounts for short-range repulsion due to Pauli exclusion, preventing atom collapse, while the $r^{-6}$ term represents weak attraction from interactions between permanent and induced dipoles such as the London dispersions.

This subsection and Table~\ref{tab:mmrep} have surveyed only the simplest and most traditional MM force fields, the efforts to enhance the expressiveness via additional terms are discussed in Section~\ref{subsec:functional-form}.

\begin{table*}[htbp]
    \centering
    \resizebox{\textwidth}{!}{%
    \begin{tabular}{c c c}
    \hline
    \textbf{Potential} & \textbf{Expression} & \textbf{Description}\\
    \hline

    \multicolumn{3}{c}{\textbf{Bond}} \\
    Harmonic potential & 
    \( U(r) = K(r - r_0)^2 \) &
    \makecell{\(K\): spring constant} \\
    Morse potential~\cite{morse1929} & 
    \( U(r) = D \left( 1 - e^{-\alpha(r - r_0)} \right)^2 \) &
    \makecell{\(D\): well depth, \(\alpha\): width of the potential}  \\
\hline \\

    \multicolumn{3}{c}{\textbf{Angle}} \\
    Harmonic potential &
    \( U(\theta) = K(\theta - \theta_0)^2 \) &
    \makecell{\(K\): bending constant} \\
    CHARMM potential~\cite{vanommeslaeghe2010charmm} & 
    \( U(\theta) = K(\theta - \theta_0)^2 + K_{UB}(r - r_{UB})^2 \) &
    \makecell{\(K_{UB}\): constant for Urey-Bradley term \\ \(r_{UB}\): equilibrium bond length between 1st and 3rd atoms} \\
    Cosine squared potential~\cite{mayo1990dreiding} &
    \( U(\theta) = \frac{1}{2} K [\cos(\theta) - \cos(\theta_0)]^2 \) &
    \makecell{\(K\): bending constant} \\
\hline \\

    \multicolumn{3}{c}{\textbf{Torsion}} \\
    Cosine potential~\cite{} &
    \( U(\phi) = K \left[ 1 + \cos(n\phi - \phi_0) \right] \) &
    \makecell{\(K\): bending constant \\ \(n\): multiplicity representing the periodicity} \\
    OPLS potential~\cite{robertson2015improved} &
    \makecell{\( U(\phi) = K_1 \left[ 1 + \cos(\phi) \right] + \frac{1}{2} K_2 \left[ 1 - \cos(2\phi) \right] \) \\
    \( + \frac{1}{2} K_3 \left[ 1 + \cos(3\phi) \right] + \frac{1}{2} K_4 \left[ 1 - \cos(4\phi) \right] \)} &
    \(K_1, K_2, K_3, K_4\): bending constants \\
   \hline \\
    
    \multicolumn{3}{c}{\textbf{van der Waals}}\\
    Lennard-Jones (12-6) potential~\cite{jones1924} & 
    \( U(r) = 4\epsilon \left[ \left( \frac{\sigma}{r} \right)^{12} - \left( \frac{\sigma}{r} \right)^{6} \right] \) &
    \(\epsilon\): well depth, \(\sigma\): distance when \( U(r) = 0 \) \\
    Buckingham potential~\cite{buckingham1938, chung2024transferability} & 
    \( U(r) = A e^{-\frac{r}{\rho}} - \frac{C}{r^6} \) &
    \makecell{\(A\) and \(C\): well depth, \(\rho\): width of the potential}  \\
    Morse potential~\cite{morse1929, hart1992van, yang2018combination} & 
    \( U(r) = D \left[ e^{-2\alpha(r - r_0)} - 2e^{-\alpha(r - r_0)} \right] \) &
    \makecell{\(D\): well depth, \(\alpha\): reciprocal length} \\ 
    Double exponential potential~\cite{wu2019double, man2021determination, horton2023transferable} &
    \( U(r) = \epsilon \left[ \frac{\beta e^{\alpha}}{\alpha - \beta} \exp \left( -\alpha \frac{r}{r_{0}} \right) - \frac{\alpha e^{\beta}}{\alpha - \beta} \exp \left( -\beta \frac{r}{r_{0}} \right) \right] \) &
    \makecell{\(\epsilon\): well depth \\ \(\alpha\) and \(\beta\): steepness of the repulsive and \\attractive interactions, respectively} \\
    \hline
    \end{tabular}%
    }
    \label{tab:mmrep}
    \caption{Representative intramolecular and intermolecular potentials~\cite{abraham_2024_11148638}. $r_0, \theta_0$ and $\phi_0$ represents equilibrium bond distance, bond angle and dihedral angle, respectively.}
\end{table*}

\section{The toolbox of composable operations for machine learning force fields.}

\label{sec:toolbox}

While there has been a plethora of previous work focused on developing and applying system-specific MLFFs to homogeneous media and benchmarking various MLFF architectures on datasets of small biologically relevant organic molecules\cite{chmiela2017machine,chmiela2022accurateglobalmachinelearning,kovacs2023evaluation,musaelian2022learning,schutt2018schnet,Unke_2019}, there have been relatively few applications to date employing MLFFs in lieu of traditional MM models for extended MD simulations of large biomolecular systems.

Part of the challenge in developing MLFFs for large biomolecular systems stems from the computational challenge of constructing datasets of \textit{ab initio} energies and forces for these large heterogeneous systems. Unlike homogeneous systems such as liquid water where one can simply train a MLFF on smaller periodic boxes or clusters of water and readily apply the trained MLFF to simulations for larger periodic boxes\cite{Chen2023dataefficient}, the chemical heterogeneity and the importance of long-range interactions in proteins\cite{Rossi2015stability} renders their decomposition into smaller, computationally more tractable training structures less straightforward. Alternatively, one could simply construct datasets encompassing the entirety of large biomolecular systems using energies and forces obtained from QM/MM evaluations. However, in addition to the QM/MM evaluations still potentially being expensive depending on the target QM method and size of the QM region, expending training resources on such large structures where most of the interactions are probably described purely at the MM-level seems wasteful.

Recently, several MLFFs trained to diverse datasets encompassing tens of thousands molecules (See Table~\ref{tab:data} and Sec.~\ref{sec:data}) have demonstrated the potential of bottom-up approaches to developing general-purpose MLFFs that can be readily applied to large biomolecules not wholly represented in the respective training sets\cite{Yao2018TensorMol,smith2017ani,smith2020ani,kovács2023maceoff23transferablemachinelearning,Ple2023forcefield,Anstine2023AIMNET2,kozinsky2023scaling,Unke2024biomolecular}. Some of the earlier general-purpose MLFFs were limited in applicability due to their training set and model architectures not explicitly accounting for charged species\cite{smith2017ani,smith2020ani}, but were still usable for conducting stable MD simulations for smaller protein systems\cite{Cheng2022building} or in mixed-level ML/MM simulations\cite{Lahey2020-tt,Rufa2020-na,Galvelis2023-ks,Zariquiey2024enhancing,Inizan2023scalable,emle}. However recently, general-purpose MLFFs based on the MACE\cite{batatia2023mace,kovács2023maceoff23transferablemachinelearning} and GEM\cite{kozinsky2023scaling} models have been successfully used to generate MD trajectories of the solvated crambin protein (18,000-25,000 atoms) for which the simulated THz-region vibrational modes, characteristic of slow collective protein motions, seem to show better agreement with the experimental spectrum than what is obtained using AmberFF\cite{kovács2023maceoff23transferablemachinelearning}.
Perhaps the largest demonstration to date of applying a general-purpose MLFF is the use of an Allegro~\cite{musaelian2022learning} model, trained to the SPICE dataset\cite{eastman2023spice}, for an all-atom MD simulation of the solvated HIV capsid consisting of 44-million atoms that achieves 8.7 timesteps/s when employing 5120 A100 GPUs\cite{kozinsky2023scaling}.

On the other hand, system-specific bottom-up approaches leveraging molecular fragmentation schemes, specifically electrostatically embedded generalized molecular fractionation with conjugate caps (EE-GMFCC)\cite{Zhang2003molecular,He2006generalized,Wang2013electrostatically}, generalized energy-based fragmentation (GEBF)\cite{Li2007generalized}, and residue-based systematic molecular fragmentation (rSMF)\cite{Wang2019toward}, have also been used to develop datasets to train MLFFs for specific proteins\cite{Wang2020combining,10.1093/bib/bbab158,Han2022inductive,Cheng2022building,Wang2019toward}. As a proof of concept, \citet{Cheng2022building} demonstrate that the GEBF method can be applied to specific proteins (1XQ8, 1013 atoms) to create relatively small datasets of fragments (5020 configurations across 65 fragments) that can be used to train accurate MLFFs for MD simulations. Unsurprisingly, their MLFFs trained to this dataset give more accurate energy and force predictions, as compared to the reference method (DFT using the $\omega$B97X-D functional), for the target protein system than the general-purpose ANI-1x\cite{smith2017ani}. 


%
%

Here, we briefly review the composable operations frequently used in the construction of MLFF in the lens of the desired properties discussed in Section~\ref{sec:desiderata}.

\paragraph{Atomistic decomposition of energy. }
Most, if not all MLFF models decompose the total energy into a sum of \textit{per-atom} energies:
\begin{equation}
\label{eq:atomistic}
U = \sum\limits_i U_i,
\end{equation}
where each $U_i$ is a function of the local atomic environments within a spatial cutoff radius.
Although widely adopted in MLFFs, this approach lacks a rigorous physical foundation, as the concept of per-atom energy is not well-defined in many-body quantum systems.
Nonetheless, this treatment is consistent with the graph-level readout that graph neural networks (GNNs) use for graph-level regression and is trivial to implement in machine learning programs.

\paragraph{Equivariant and invariant features.}
It has been outlined in Section~\ref{sec:desiderata} that an ideal force field $\hat{U}(\mathbf{x})$ should be an $E(3)$-invariant function of $\mathbf{x}$.
It has been illustrated in \citet{Unke_2019, thölke2022torchmdnet, schütt2021equivariant, wang2023spatial}, however, that intermediate \textit{equivariant} representations can boost the expressiveness and performance of invariant models.
Concretely, w.r.t. a group $G$, an invariant function $y = f(\mathbf{x})$ can be constructed as the product of an invariant function $g_\mathtt{I}$ and an equivariant function $g_\mathtt{E}$ as:
\begin{equation}
y = f(\mathbf{x})
= g_\mathtt{I}(
g_\mathtt{E}(\mathbf{x})
)
\end{equation}
The equivariant-invariant mapping $g_\mathtt{I}$ is also called a \textit{scalarization}.
In practice, a modern MLFF usually keeps track of both invariant and equivariant features and updates them simultaneously.
So the functional signature of a $G$-equivariant MLFF layer operating on both invariant $h \in \mathcal{H}$ and equivariant $\mathbf{x} \in \mathcal{X}$ (w.r.t. the same space) features $f: \mathcal{H} \oplus \mathcal{X} \rightarrow \mathcal{H} \oplus \mathcal{X} $ can be written as:
\begin{equation}
\label{eq:two-inputs}
\mathbf{x}', h' = f(\mathbf{x}, h), 
\end{equation}
where we have:
\begin{equation}
T_g(\mathbf{x}'), h' = f(T_g (\mathbf{x}), h),
\forall T_g \in G.
\end{equation}

\paragraph{Message passing. }
Most \textit{spatial} GNNs~\cite{duvenaud2015convolutional, kipf2016gcn, gilmer2017neural, xu2018powerful, battaglia2018relational, jain2018topology, wu2019simplifying, wang2019deep, wang2019dynamic, joshi2023expressive, wang2021stochasticaggregationgraphneural} on topological graphs (without geometry features) adopt a message passing framework.
Following the framework from \citet{xu2018powerful} and \citet{battaglia2018relational}, the $k$-th layer of a GNN could be written as two steps---\textit{neighborhood aggregation}:
\begin{equation}
\label{eq:agg}
a_v^{(k)} = \rho^{(k)}
\big(
h_u^{(k-1)}, u \in \mathcal{N}(v)
\big),
\end{equation}
and \textit{node update}:
\begin{equation}
\label{eq:update}
h_v^{(k)} = \phi^{(k)}(h_v^{(k-1)}, a_v^{(k)}),
\end{equation}
where $h_v^{k}$ is the feature of node $v$ at $k$-th layer, $h_v^{0} = \mathbf{x}_v$ and $\mathcal{N}(\cdot)$ denotes the operation to return the multiset of neighbors of a node. 
More concisely, omitting the nonlinear transformation step $\phi$ ubiquitous in all neural models, and assuming a convolutional \textit{aggregate} function, $\rho = \operatorname{SUM}$ or $\rho = \operatorname{MEAN}$, a graph neural network layer is characterized by the aggregation/convolution operation that pools representations from neighboring nodes, forming an intermediary representation $\mathbf{X}'$, which on a global level, with activation function $\sigma$ and weights $W$, can be written as:
\begin{equation}
\label{eq:gnn-master}
    \mathbf{X}' = \sigma(\hat{A}\mathbf{X}W)
\end{equation}
The difference among GNN architectures, apart from the subsequent treatment of the resulting intermediate representation $\mathbf{X}'$, typically amounts to the choices of transformations ($\hat{A}$) of the original adjacency matrix ($A$)---the normalized Laplacian for graph convolutional networks (GCN)~\cite{kipf2016gcn}, a learned, sparse stochastic matrix for graph attention networks (GAT)~\cite{velickovic2018graph}, powers of the graph Laplacian for simplifying graph networks (SGN)~\cite{DBLP:journals/corr/abs-1902-07153}, and the matrix exponential thereof for graph neural diffusion (GRAND)~\cite{DBLP:journals/corr/abs-2106-10934}. 
To expand these frameworks to incorporate geometry information, the layers to incorporate coordinates surveyed in this section can be plugged into Equation~\ref{eq:agg} and Equation~\ref{eq:update}.
We also note that the convolution operator in Equation~\ref{eq:gnn-master} is the root of a plethora of performance pathologies including over-smoothing~\cite{cai2020note, rusch2023survey}, over-squashing~\cite{DBLP:journals/corr/abs-2006-05205, topping2022understanding}, and limited expressiveness~\cite{corso2020principal, DBLP:journals/corr/abs-2002-06157, xu2018powerful}, and alternative forms of GNNs~\cite{wang2024nonconvolutionalgraphneuralnetworks} might be needed to address them to make them more expressive and robust.

\paragraph{Gaussian smearing. }
Distances among particles are arguably the simplest and most crucial invariant feature---with a distance matrix, the coordinates of the particles can be reconstructed modulo the $E(3)$-equivariance, and all invariant functions can be approximated arbitrarily well.
Nevertheless, to blindly throw the distances into a neural network yields only highly correlated representations~\cite{schütt2017schnet}, as the detailed change in the distances is unlikely to be reflected in the activation of a large neural network.
Since \citet{PhysRevLett.98.146401}, the Gaussian-smeared distances have been used widely as a radial, invariant feature.
The most generic form of Gaussian smearing resembles the radial basis function (RBF) kernel:
\begin{equation}
\label{eq:rbf}
K(x, x') = \exp\left(-\frac{\|x - x'\|^2}{2\sigma^2}\right).
\end{equation}
By carefully choosing a set of evenly spaced $x'_0, x'_1, \ldots, x'_i, \ldots x'_k$, Equation~\ref{eq:rbf} yields a $k$-dimensional vector with maximal convoluted signal at $x'_i$, resolution defined by $x'_{i+1} - x'_{i}$, and sensitivity controlled by $\sigma$.

\paragraph{Angular symmetry function. }
Also proposed in \citet{PhysRevLett.98.146401}, the angular symmetry function takes the angles among triplets of atoms (which can be calculated using Equation~\ref{eq:geometry}) directly in a neural network.
This term closely resembles how MM encodes angular environment, albeit with a more expressive function.
Many popular MLFFs, most notably \citet{Smith_2017}, incorporate this feature locally to account for angular environments.

\paragraph{Dot product scalarization. }
If we center our view around each particle and do not consider the translation invariance, the group we consider would be $O(3)$ rather than $E(3)$ and all $O(3)$-invariant functions can be universally approximated using the dot products among the inputs.
\begin{lemma}[First Fundamental Theorem~\cite{villar2023scalars} for $O(n)$]
\label{lemma:first-fundamental}
If $f$ is an $O(n)$-invariant scalar function of vector inputs $v_1, \ldots, v_n \in \mathbb{R}^{D}$, then $f(v_1, v_2, \ldots, v_n)$ can be written as a function of only the scalar products of the $v_i$.
That is, there is a function $g(\cdot)$ such that
\begin{equation}
f(v_1, v_2, \ldots, v_n)
= g(V^\intercal V)
= g((v_i^\intercal v_j)^{n}_{i, j = 1})
\end{equation}
\end{lemma}
This lemma is the theoretical cornerstone of a number of locally universal MLFF models~\cite{Unke_2019, thölke2022torchmdnet, schütt2021equivariant, wang2023spatial}.
It is worth mentioning, however, that function $g$ in Lemma~\ref{lemma:first-fundamental} is not necessarily permutation invariant, and extra care is needed to design $O(3)$-invariant, permutationally invariant, and universal functions.
\citet{wang2023spatial}, for example, neurally parametrizes a series of edge vectors prior to the dot product.

\paragraph{Linear combination of equivariant features.}
Ignoring the translational equivariance again and considering only the $O(3)$ (or $O(n)$) group, the linear combination of equivariant features is naturally equivariant, and the linear combination of equivariant \textit{maps} ($\{ f_i \}$) 
\begin{equation}
F(\mathbf{x}) = \sum\limits_{i}\lambda_i f_i(\mathbf{x})
\end{equation}
where $\lambda_i$ are constants, are also equivariant.
In practice, this summation can be implemented as a linear transform, or a single-layer neural network without biases or activation, operated on the last dimension of a $\mathbb{R}^{N, 3, D}$ vector representation.
Moreover, the coefficients $\{ \lambda_i \}$ can also be calculated from the invariant representations.

\paragraph{Spherical harmonics.}
\citet{batzner2021se, musaelian2022learning, batatia2023mace, DBLP:journals/corr/abs-1802-08219} construct $SO(3)$-equivariant (removing both the translation and reflection transformations from $E(3)$) convolution filters to be products of radial (invariant) functions and spherical harmonics,
\begin{equation}
\label{eq:spherical}
F(\vec{r}) = R(r)Y_m^{(l)}(\vec{r}),
\end{equation}
which can be contracted with the Clebsch-Gordan coefficients to form higher-order tensor products.
Currently, the class of models using these representations make up the most accurate MLFFs, although it has been outlined in Lemma~\ref{lemma:first-fundamental} that they are not required for universality.

\section{Best practices and pitfalls.}
We have been insofar discussing the functional form of force fields $\hat{U}(\mathbf{x}; h, \Phi)$, but in order to design a practical $\hat{U}$, one needs to find a set of optimal parameters by maximizing the likelihood of some particular force and energy data.
In this section, we briefly review the popular datasets used in the curation of MM and ML force fields and the best practices for training and evaluation.

\subsection{Datasets}
\label{sec:data}

\begin{table*}[htbp]
    \centering
    \begin{tabular}{c c c c c c c c}
    \hline
    Dataset 
    & Elements 
    & Chemical spaces 
    & \# Molecules 
    & \# Conformers 
    & Level of theory 
    & Sampling & \\\hline
    MD17~\cite{chmiela2017machine, christensen2020rolegradientsmachinelearning} 
    & C,H,O 
    & \makecell{Small molecules \\ up to 21 atoms}
    & 10 
    & 2.7m
    & DFT
    & Path-integral MD \\
    MD22~\cite{chmiela2022accurateglobalmachinelearning} 
    & C, H, O, N
    & \makecell{Small molecule \\w/ 42 $\sim$ 270 atoms}
    & 7
    & 22k
    & PBE+MBD~\cite{perdew1996generalized, tkatchenko2012accurate} 
    & MD\\
    QM9~\cite{ramakrishnan2014quantum} 
    & C, H, O, N, F
    & Small molecule
    & 134k
    & 134k
    & B3LYP/6-31G(2df,p)
    & Minimized\\

    ANI-1~\cite{smith2017anidata}
    & C, N, O, F
    & \makecell{Small molecules \\ up to 11 heavy atoms}
    & 57,462
    & 24,687,809
    & DFT
    & \makecell{DFT-optimized \\ + normal mode}\\

    ANI-2~\cite{smith2020ani}
    & H, C, N, O, S, F, Cl
    & small molecules
    & 13,405
    & 4,695,707
    & wB97X/631Gd
    & \makecell{dimer, normal mode,\\ \\torsion sampling, \\and active learning}\\

    OrbNet Denali~\cite{Christensen2021, Christensen_2021}
    & \makecell{H, Li, B, \\ C, N, O, F, \\Na, Mg, Si, P, S, \\ Cl, K, Ca, Br, I}
    & organic molecules
    & DFT
    & 15,000
    & 2.3 million
    & normal mode + MD\\
    
    SPICE~\cite{eastman2023spice} 
    & \makecell{Li, C, N, O, F, Na, Mg, \\ P, S, Cl, K, Ca, Br, I}
    & \makecell{Small molecules \\peptides, ion pairs}
    & 19k
    & 1.1m
    & \makecell{$\omega$B97M-D3(BJ)\\/def2-TZVPPD}
    & MD + Minimization\\

    \hline
    \end{tabular}
    \caption{Popular datasets used to curate force fields.
    For a more comprehensive list, see \citet{ullah2024molecular}.
    }
    \label{tab:data}
\end{table*}

\paragraph{A note on the QM target.}
There has been little consensus, when it comes to biomolecular applications, regarding which QM levels of theory might correspond best to experimental measurements~\cite{cavasotto2020binding}.
As a result, popular datasets (see Table~\ref{tab:data}) are curated with various levels of theory and sampling strategies, making the merging of the data difficult, demanding meta-learning solutions~\cite{allen2023learningtogetherfoundationalmodels}.
Before further evidence would emerge, if the field were to agree on a single level of theory for a community-wide effort to push for high-quality, high-volume data for MLFF training, cheaper targets might be more appealing.

When developing novel MLFFs, it is also worth reminding ourselves that all we are trying to fit is a known function that can be solved analytically.
One can view the QM energy function not as a reservoir of data but a \textit{surrogate}, from which repeated acquisition is possible.
Active learning techniques, such as \citet{smith2018less, wang2020active, schwalbe2021differentiable}, present a useful avenue to gather data in a rationally parsimonious fashion.

\paragraph{Chemical diversity and conformational diversity.}
Conceptually, to accurately fit energies and forces on \textit{all} chemical spaces is no different than having the MLFF model able to solve the Schrödinger's equation, which seems impossible, judging from the \textit{no free lunch} theorem~\cite{wolpert1996lack}.
To this end, QM datasets are always curated with biases, in terms of the coverage on chemical spaces and conformational landscape.

Within the chemical space of (bio)organic molecules, MD17~\cite{chmiela2017} is among the most popular datasets used for MLFF benchmark.
Consisting of DFT calculations of 10 small organic molecules (benzene, uracil, naphthalene, aspirin, salicylic acid, malonaldehyde, ethanol, toluene, paracetamol, and azobenzene) with fewer than 1 million snapshots each, it is limited in chemical diversity.
\citet{christensen2020role} further revised this dataset to reduce the noise.
There are also high-quality QM datasets focusing on single molecules such as 3BPA~\cite{kovacs2021linear}, and AcAc~\cite{batatia2022design}.
ISO17~\cite{schütt2017schnet}, on the other hand, is slightly more chemically diverse as it samples $129$ isomers of \ce{C7O2H10} with $5000$ snapshots each.
In contrast, QM9~\cite{ramakrishnan2014quantum}, like its predecessors, is rich in chemical diversity but not conformational diversity---it contains more than $134$k small molecules, although all in low-energy state.

Moving on to larger datasets emphasizing \textit{utility} beyond just proof-of-concept, ANI1~\cite{smith2017anidata} and ANI2~\cite{devereux2020extending} dataset, containing 20 million off-equilibrium snapshots, is among the most popular datasets that are simultaneously chemically and conformationally diverse, albeit no QM forces were annotated, which would be more information-rich than energies.
SPICE~\cite{eastman2023spice, eastman_2024_10975225} also displays a vast chemical and conformational diversity on $2$ million conformations of small molecules and peptides, with forces annotated.
The SPICE~\cite{eastman2023spice} dataset, along with many diverse datasets for (bio)molecules, has been generated on the QCFractal platform~\cite{smith2021molssi}, which is being used to actively run QM calculations to curate the next generation of datasets for MLFF fitting.

Another community-driven project, ColabFit~\cite{vita2023colabfitexchangeopenaccessdatasets} Exchange curates an open-source, diverse database containing a large collection of systematically organized datasets from multiple chemistry/materials domains that are especially designed for ML atomistic model development (providing physical units standardization, a unified data standard, integrated data loaders, etc.). 
To date, the database contains 372 datasets containing more than 180M atomic configurations and 500M properties, spanning over 100 thousand different chemistries, and it is constantly expanding.

In principle, topology-free MLFFs are more naturally suited for simulating chemical bond-breaking and forming events than standard FFs, and would enable the simulation of fundamental biochemical processes like enzymatic reactions for which \textit{ab initio} accuracy and FF efficiencies would ideally be employed. However, in practice, the curation of datasets to train MLFFs to accurately model chemical reactions presents additional challenges\cite{Yang2024machine}, the same age-old challenges associated with sampling rare events, as compared to developing models for equilibrium sampling of unreactive systems. In order to efficiently sample higher-energy training configurations representative of reaction pathways, various techniques have been employed ranging from simply running unbiased MD simulations at elevated temperatures\cite{Rivero2019reactive,zhang2024exploring} and identifying minimum energy paths\cite{Gastegger2015high,Schreiner2022transition1x,Young2022reaction} to other approaches incorporating biased MD simulations leveraging enhanced sampling algorithms in active learning protocols\cite{Pan2021machine,Devergne2022combining,Yang2022using,Benayad2024prebiotic}.

Lastly, we note that training on condensed-phase properties has only been very recently~\cite{magduau2023machine} possible.
MM force fields are, on the other hand, usually successful in faithfully reproducing these properties~\cite{boothroyd2022open}.
These largely depend upon the \textit{inter}molecular interactions, whereas long-range interactions are traditionally neglected in MLFFs (prior to SPICE~\cite{eastman2023spice}, the datasets used to curate MLFFs only include intramolecular interactions).

\subsection{Training and evaluation}
\label{sec:train}
We observe that, in recent literature on MLFFs, the community is converging on a set of practices for efficient training and fair comparison of MLFF models.
Certain practices, such as extremely small batch size, exponentially decaying learning rate, and parsimonious use of normalization, are common in the curation of highly performant MLFFs.
Since the forces and energies, and thereby force errors $\mathcal{L}_F$ and energy errors $\mathcal{L}_U$, for instance measured in mean-squared-error, are of different units, one has to apply a set of perhaps physically meaningless constants to combine them as the loss function $\mathcal{L}$:
\begin{gather}
\mathcal{L}_U = ||\hat{U} - U ||^2, \\
\mathcal{L}_F = \frac{1}{3N} \sum\limits_{i=1}^N
\sum\limits_{\alpha=1}^{3} || -\frac{\partial \hat{U}}{\partial r_{i, \alpha}} - F_{i, \alpha}||^2 \label{eq:mse_f},\\
\mathcal{L} = \lambda_U \mathcal{L}_U + \lambda_F \mathcal{L}_F.
\end{gather}
Empirically, when using atomic units, $\mathcal{L}_F / \mathcal{L}_U$ within the range $100\sim 1000$ usually yields the best results.

It is worth reminding that, while the MSE (Equation~\ref{eq:mse_f}) and RMSE error on forces are $E(3)$-invariant, the MAE loss on forces is \textit{not}, and is dependent upon the choice of the coordinate systems.
\begin{equation}
\mathcal{L}_\mathtt{F, MAE} = \frac{1}{3N}\sum\limits_{i=1}^N\sum\limits_{\alpha=1}^{3} | -\frac{\partial \hat{U}}{\partial r_{i, \alpha}} - F_{i, \alpha}|
\end{equation}
In a sense, this error has a bias to favor the conformations more aligned with the axes.
It is alarming to see that this biased and arguably erroneous metric has been used widely in both the training and evaluation stages of MLFF models incorporating force matching.

The aforementioned error is a typical example of how error-prone the implementation of MLFF modules is---some intuitively benign operations might break the symmetry without catching the eyes of a seasoned engineer-researcher.
Thus, we recommend that an equivariance/invariance unit test be included in all modules of MLFF implementation.
An example test suite for a function $f$ that works on both $SE(3)$ invariant ($h$) and equivariant ($x$) representations (See Equation~\ref{eq:two-inputs}) can be implemented in 5 lines, using NumPy~\cite{Harris_2020}, for example:

\begin{python}
def test_equivariant_and_invariance(f, h, x):
    # random translation
    import numpy as np
    T = np.random.randn(1, 3)

    # random rotation
    R = np.linalg.qr(
        np.random.randn(3, 3))[0]

    # random transformation
    F = lambda x: x @ R + T

    # assert h changed
    # and x transformed accordingly
    h, x = f(h, x); h1, x1 = f(h, F(x))
    assert (
        np.allclose(h, h1) 
        & np.allclose(F(x), x1)
    )
    
\end{python}




\section{Making MM more accurate}
\subsection{Functional forms: more, but only slightly more, than harmonics and Fourier series.}
\label{subsec:functional-form}
The choice and design of functional forms (see Table~\ref{tab:mmrep}) themselves limit the flexibility and expressiveness of the force field. 
For instance, the Lennard-Jones 12-6 potential, described in Equation~\ref{eq:u_mm}, was developed decades ago and has been widely adopted since, despite the existence of alternative approaches~\cite{halgren1992representation, man2021determination}.
Enriching the complexity of the MM functional form has long been an area of intensive research. 
Notably, Class II MM force fields~\cite{maple1994derivation, hwang1994derivation-2,maple1994derivation-3} replace the harmonic terms in Equation~\ref{eq:u_mm} with more much more flexible terms such as:
\begin{itemize}
    \item \textit{Higher order polynomials:}
    The harmonic terms for bonds and angles can be rewritten to incorporate higher $k$-order polynomials in the form of (reusing the notation from Equation~\ref{eq:u_mm}):
    \begin{equation}
        \sum\limits_\mathtt{bond} \sum\limits_k K_k(r_{ij} - r_k)^2; 
        \sum\limits_\mathtt{angle} \sum\limits_k K_k(
        \theta_{ij} - \theta_k)^2.
    \end{equation}
    Since the geometry of a system is uniquely defined by the inter-atomic distances, given sufficiently high-order polynomials and dense enough bond connections, this functional form can be made universal.

    \item \textit{Coupling terms:}
    Terms like bond displacement, angle displacement, and Fourier series for (proper and improper) torsions can be combined multiplicatively to form bond-bond coupling (showing here as an example), 
    \begin{eqnarray}
    \sum\limits_\mathtt{bond}\sum\limits_\mathtt{bond}
    K_\mathtt{bond, bond} (r_{i_0 j_0} - r_0) (r_{i_1 j_1} - r_1),
    \end{eqnarray}
    angle-angle coupling, bond-angle coupling, bond-torsion coupling, angle-torsion coupling, or torsion-angle-angle coupling.    
\end{itemize}

In a similar spirit, \citet{Xie_2023} constructs ultra-fast machine learning potentials using B-splines and parametrizes using machine learning approaches.
On a simple test system, this force field has a favorable tradeoff between speed and accuracy and is guaranteed to be smooth.

On the other hand, polarizable force fields, such as Drude~\cite{lemkul2016emperical} and AMOEBA force fields~\cite{shi2013polarizable}, incorporate the ability to dynamically adjust the distribution of atomic charges in response to the local electrostatic environment and generally involve a self-consistent calculation of induced dipoles.
These can be augmented with neural networks to account for missing contributions~\cite{illarionov2023mlip}.
Reactive force fields, such as ReaxFF~\cite{duin2001reaxff, kaymak2022jax-reaxff}, can handle bond breaking and forming during simulations by dynamically updating bond orders, based upon interatomic distances, for every simulated MD frame and making the relevant energy terms dependent on those bond orders.
Empirical valence bond (EVB) models~\cite{warshel1980empirical} also enable the use of standard MM force fields for reactive simulations\cite{lobaugh1996quantum,sagnella1998empirical,schmitt1998multistate,aqvist1993simulation}, but require the parametrization of a coupling term between the reactant and product states for a particular system. These have traditionally been system-specific, but recent work has focused on developing transferable parametrizations of EVB models across different reactive systems\cite{allen2024toward}.
However, improving force field accuracy by introducing more expressive and additional functional terms comes with increased computational costs, which may not justify the trade-off between accuracy and speed, and significantly increases the complexity of force field parametrization.

\subsection{Parametrization: from engineer-years to GPU-days}
\label{subsec:parametrization}
Another challenge in developing a reliable, robust and extensible MM force field is the parametrization scheme---the assignment of the parameter set $\Phi_\mathtt{FF}$---which must ensure comprehensive chemical coverage across the broad and heterogeneous chemical space relevant to biomolecular systems.

The determination of $\Phi_\mathtt{FF}$ has traditionally been reliant on a human labor-intensive, inflexible, and inextensible rule-based scheme named \textit{atom typing}---it classifies atoms into discrete categories representing distinct chemical environments. This classification enables MM parameters to be subsequently assigned from a tabulated table of relevant atomic, bond, angle, and torsion parameters. 
For example, in the case of small molecules, atom types are determined by the attributes of the atom, such as element identity, hybridization, and aromaticity, as well as the attributes of the neighboring atoms and their connected bonds, and the number of neighboring atoms. 
For amino acids, the atom types are traditionally assigned according to the residues. 
Most of these atom types have a receptive field of two or three bonds, and chemical motifs outside this receptive field are generally not realized.
After the atom types are determined, the bond, angle, and torsion types are determined simply by composing (using the $\operatorname{AND}$ operation and dictionary look up) atom types, and as a result, $K$ atom types can naively lead to $K^4$ torsion types without simplification.
The van der Waals interactions, on the other hand, are usually described with Lennard-Jones 12--6 potentials using the Lorentz-Berthelot~\cite{delhommelle2001inadequacy} combining rules to determine $\sigma$ and $\epsilon$ between different atom types.

The force field parameters can be further optimized in a systematic manner using ensemble reweighting method~\cite{wang2013systematic, wang2014building, thaler2021learning} and machine-learning methods~\cite{befort2021machine, wang2023dmff}. However co-optimizing the discrete chemical perception defined by the rule-based atom types and continuous force field parameters remains intractable.
In general, the force field accuracy is constrained by the resolution of chemical perception. 
Efforts to improve the accuracy by increasing the number of atom types lead to a combinatorial explosion of required types. 

Although there are efforts to automate the development and parametrization process~\cite{koziara2014testing, betz2015paramfit, harder2016opls3, horton2022open, kumar2024ffparam}, human expertise remains essential, introducing challenges in adjusting existing parameters to accommodate new ones, particularly when extending the force field to new chemical domains of interest.
In addition, biomolecular systems are inherently heterogeneous, making MM force field optimization challenging. 
The popular AmberTools~23 package, for instance, combines independently developed force fields for chemical subspaces including proteins~\cite{ff14SB}, DNA~\cite{OL15,OL15full}, RNA~\cite{OL3}, water~\cite{jorgensen1983comparison,TIP4P-EW,OPC}, monovalent~\cite{joung2008determination,joung2009molecular} and divalent~\cite{li2013rational,li2014taking,li2015parameterization} counterions, lipids~\cite{dickson2022lipid21}, carbohydrates~\cite{kirschner2008glycam06}, glycoconjugates~\cite{demarco2009atomic,demarco2010presentation}, small molecules~\cite{wang2004development,wang2006automatic}, post-translational modifications~\cite{khoury2013forcefield_ptm}, and nucleic acid modifications~\cite{aduri2007rna}---crystallized from more than 100 engineer-years of effort.
Nevertheless, there is no guarantee that the optimized solution for each class will constitute the globally optimized solution.

\paragraph{Substructure pattern matching}
Substructure pattern matching approaches~\cite{mobley2018escaping, stroet2023oframp, yesselman2012match} represent another class of force field parametrization schemes that focus more directly on chemical perception than the traditional atom-typing methods.
The Open Force Field Initiative~\cite{wang2024open}, for instance, has developed an ecosystem of toolkits~\cite{jeff_wagner_2024_10613658} and force field releases~\cite{boothroyd2023development} that use standard SMARTS-based chemical substructure queries~\cite{mobley2018escaping} to assign entire sets of valence parameters (atoms, bonds, angles, torsions) in a hierarchical manner. 
These approaches help mitigate the combinatorial explosion of parameters and significantly reduce atom type redundancy while maintaining the force field accuracy.
For example, the latest Sage force field (openff-2.2.0~\cite{mcisaac_2024_10995191}) from the Open Force Field Initiative contains 187 torsion parameters, approximately 800 times fewer than the OPLS3e force field~\cite{roos2019opls3e}, which relies on atom-typing.
Furthermore, the reduced complexity of substructure-based approaches facilitates the automated fitting of parameters for specific force fields, such as through the use of ForceBalance~\cite{wang2014building}.
However, determining and refining substructure patterns for more reliable force fields still requires a human-in-the-loop approach, combining human expertise with automated~\cite{gokey2023hierarchical} procedures.


\paragraph{Graph-based chemical perception: molecular topology as a graph}
Applying graph neural networks (GNNs) for more robust and extensible MM force field parametrization is another emerging area~\cite{D2SC02739A, takaba2023machinelearned, takaba2023machinelearnedARXIV, chen2024advancing, seute2024grappa, thurlemann2023regularized, wang2023dmff, lehner2023dash, wang2024espalomacharge, wang2019graphnetspartialcharge, wang2023graph}. 
For example, ~\citet{D2SC02739A, takaba2023machinelearned} have demonstrated the ability to replace traditional rule-based discrete atom-typing schemes with continuous atomic representations generated by neural networks operating directly on chemical graphs using an end-to-end differentiable framework. 
The neural network parameters are optimized directly through standard machine learning frameworks to fit quantum chemical and/or experimental data. 
These approaches enable the co-optimization of chemical perceptions, represented as continuous atom embeddings, alongside continuous force field parameters.
For example, the latest Espaloma force field~\cite{takaba2023machinelearned}, trained in less than one GPU day on a vast chemical space (comprising $17$k molecules and over $1$ million QM snapshots) consisting of small molecules, peptides, proteins, and RNAs, has shown the capability to accurately predict not only energy and forces but also, when used in MD simulations, NMR observables, and protein-ligand binding free energies. This demonstrates a promising path forward for the flexible and efficient curation of molecular mechanics (MM) force fields.
Nevertheless, Leonard-Jones parameters are \textit{not} learned in this framework, which is required to complete a force field.


\section{Making ML force fields faster}
\label{sec:makingMLfaster}
As of now, almost all the fastest ML force fields use dot-product scalarization (Lemma~\ref{lemma:first-fundamental}) to compute the energies and automatic differentiation to compute the forces.

\paragraph{Accelerating SO(3) convolutions.}
Although the spherical harmonics-based SO(3)-equivariant representations have been shown to be performant and data-efficient in constructing MLFFs~\cite{batzner2021se, musaelian2022learning, batatia2023mace, DBLP:journals/corr/abs-1802-08219}, an $L$-degree tensor convolution would require $\mathcal{O}(L^6)$ complexity.
\citet{passaro2023reducingso3convolutionsso2} addresses this issue by reducing SO(3) convolutions to SO(2) without losing information.
\citet{luo2024enablingefficientequivariantoperations} also achieves such cubic complexity by performing the convolution in the Fourier space and using Gaunt coefficient rather than the Clebsch-Gordan coefficients.
\citet{cheng2024cartesianatomicclusterexpansion}, on the other hand, performs the \textit{atomic cluster expansion}~\cite{kovacs2021linear} directly in the Cartesian coordinate system.

\paragraph{Machine learning for coarse-graining.}
Coarse-graining (CG)~\cite{smit1990computer, muller2006biological, marrink2007martini, souza2021martini} refers to the technique to group atoms into larger particles termed \textit{beads}, whose interactions are used to approximate the interaction energy among atoms.
This enables the simulation of slow, collective motions while \textit{averaging out} fast, local movements.
For $N$ atoms with coordinates $\mathbf{X} \in \mathbb{R}^{N \times 3}$ and $n$ beads $\mathbf{x} \in \mathbb{R}^{n \times 3}$, the CG operator can be written as:
\begin{equation}
\label{eq:cg}
\mathbf{x} = P\mathbf{X},
\end{equation}
where $P$ is right-stochastic ($\sum_j P_{ij} = 1$), hence the translation and rotation equivariance discussed in Section~\ref{sec:desiderata} are naturally satisfied.
While $P$ has traditionally been pre-defined and discrete, it can be made continuous and optimizable with machine learning~\cite{wang2019machine, wang2019coarse, yang2023chemically}.
Coincidentally, Equation~\ref{eq:cg} also closely resembles the linear projection step in Linformer~\cite{wang2020linformerselfattentionlinearcomplexity}, an approach to reduce the complexity of transformer~\cite{vaswani2023attentionneed} models from quadratic to linear.

\section{The path forward: what would the next generation of force fields look like?}
\label{sec:dream}

\subsection{The dilemma of topology.}
It is debatable to what degree chemical bonds (and angles and torsions) are \textit{real} and not artificial constructs.
Classical biomolecular MM force fields usually require a \textit{topology} (exceptions include \citet{gale2021universal}) and to define bond and angle energy accordingly, which is equivalent to putting a (very strong) prior on the probability density $\hat{p}$ in Equation~\ref{eq:forcefield} to restrict $\mathbf{x}$ in a limited region.
Such intense biases in a simulation ensure its stability and interpretability---clashes and distortions in the geometry will have near-zero likelihood under such formulation.
At the same time, it also prohibits the force field model reactive species and transition states.

Anecdotally, during the implementation of an MD simulation, most time and effort of a researcher is typically spent on defining the \textit{topology}---protonations and tautomers---of the system.
Domain knowledge is also crucial in setting up an MD simulation, as even a protonation state error can drastically change the entire energy landscape of a biomolecular system~\cite{gunner2020standard}.

MLFF, on the other hand, usually does not have the notion of \textit{topology} entirely (exceptions exist, such as \citet{eastman2024nutmegspicemodelsdata}, which uses the molecular topology to define a set of pre-computed partial charges but does not explicitly restrain bonds and angles).
Faithful to Equation~\ref{eq:forcefield}, it typically takes only the geometry $\mathbf{x}$ and the element identity $h$ as input without restricting $\mathbf{x}$ on any subspace.
Naturally, this mitigates the need for a carefully designed topological graph and can, in theory, simulate transition states and reactive species.
There is no guarantee, however, that the simulation will stay stable and interpretable~\cite{fu2023forces}, especially on high-energy regions.

To be topology-free can mean that the topological information needs to be re-realized every time.
One can think that each forward pass of an MLFF model entails both the \textit{topology realization stage} and the \textit{inference} stage in an MM force field.
As such, even for small perturbations on a limited conformational space, the computation always starts from scratch.
We are interested in studying whether the realization of crude topology can be cached for similar conformations.

In addition, being free of topology also means that there are no inductive biases for a model to avoid apparently unfavorable regions.

\subsection{What functional terms are here to stay?}
\textit{Dot-product scalarization.}
As examples of Lemma~\ref{lemma:first-fundamental}, all terms in Equation~\ref{eq:u_mm} can be written as functions of dot products of edge (chemical bond) vectors, as one can trivially verify.
For example, self dot product recovers the distance, and angles can be calculated as the ratio between dot products and distances.
Since it is already universal, all invariant functions approximated by spherical harmonics can also be approximated by dot product-based algorithms, and spherical harmonics-based methods only afford the model with extra (physically inspired) inductive biases so that they are more data-efficient.
Moreover, since dot product is among the most ubiquitously used operations in machine learning (for instance in attention~\cite{vaswani2023attention}), it has been most aggressively optimized even for extremely high dimensional vectors, whereas optimizing spherical harmonics operations in GPU is still a technical challenge.
We anticipate that dot product-based MLFF models will show more utility in the coming years.

\textit{Long-range interactions.}
Most, if not all, MLFF models adopt a cutoff function (Equation~\ref{eq:cutoff}) and interactions between particles more than 5 or 10 angstroms apart are all masked out.
This means that crucial interactions highly meaningful in key biomolecular processes such as protein folding, which has long been modeled using MM force fields~\cite{larson2009foldinghome}, cannot even be realized by MLFFs.
One viable approach to address this limitation would be to incorporate the van der Waals term in an MLFF model~\cite{doi:10.1021/acs.jctc.1c01021}.
A $\Delta$-learning can be carried out to learn the force and energy difference between the QM reference and the van der Waals interaction from an existing force field.
Optionally, one can adopt a similar scheme with \citet{D2SC02739A, takaba2023machinelearned} to jointly optimize the parameters of the van der Waals terms, although alternative~\cite{horton2023transferable} formulations other than the 12-6 term might be preferred, such as those with more physical meaning namely the charge equilibration models~\cite{rappe1991charge} or the Buckingham~\cite{buckingham1938} potential.

Recall the atomistic decomposition of energy (Equation~\ref{eq:atomistic})---this approach inherently assumes that the local structure is the primary determinant of interaction energy, ignoring contributions from long-range interactions.
This approximation limits the applicability of the developed models to local interactions \cite{Ko2021-yz}. 
To address this limitation, a hierarchical framework has been introduced by \citet{Ko2021-yz}.
While machine learning force fields can use Equation~\ref{eq:atomistic} without further modifications, new~\cite{Unke_2019, Ko2021-yz} generations of ML force fields perform a decomposition into short-range energy (equivalent to Equation~\ref{eq:atomistic}) and a long-range electrostatic (and, sometimes, dispersion) contribution~\cite{anstine2023machine}, written as:

\begin{equation}
U = \sum\limits_i U_i + U_{\text{lr}}
\end{equation}

Long-range interactions $U_{\text{lr}}$ typically contain two relevant contributions: electrostatic and dispersion interaction.
Electrostatic interactions are typically calculated using environment-dependent partial charges $q_i$ via a neural network coupled with a charge equilibration procedure in neural networks.
London dispersion forces represent another significant and essential component of long-range interactions. 
These have often been neglected or included through a Grimme two-body (pairwise additive) or three-body interactions and environment-dependent correction, though there are notable exceptions \cite{Westermayr2022-cj, Tu2023-pg}.

\subsection{Mixing MM with ML potential}
\label{sec:mmml}

Apart from incorporating pairwise functional forms for the long-range interactions, we can also further mix components of MM and ML potential models.

\paragraph{Mixing energy functions}
ML/MM approaches, in direct analogy to QM/MM, can in principle help to close this timescale divide, and several recent works have demonstrated proof of principle on how such approaches could be employed to more efficiently model large biomolecular systems\cite{Lahey2020-tt,Rufa2020-na,Boselt2021machine,Galvelis2023-ks,Inizan2023scalable,Zariquiey2024enhancing,emle}(See Sec.~\ref{sec:mmml}). 
Recognizing the speed difference between MM and ML force fields and assuming that its ratio will remain near constant in the next generations of ML and MM force fields (which is a realistic assumption) leads to the question: is it necessary to simulate the entire molecular system with an ML force field when the region involved in the event of interest is often limited to a small subsystem (reaction center, interactions between specific amino acids and a small molecule).
This is further motivated by the insight that generating an MM force field that generalizes well is challenging.

The `subtractive' scheme or `mechanical' embedding is the simplest and easiest to implement approach \cite{Senn2009-wp, Lahey2020-tt, cole2020machine}. 
In such a formalism, the ML force field describes the intramolecular energetics $V_{ML}$ of the ligand $\vec{r}_{L}$ and the MM force field is responsible for the interaction between the ligand and the environment $E$.   

\begin{equation}
    V_{MM/ML} = V_{MM}(\vec{r}_{E}, \vec{r}_{L}) + V_{ML}(\vec{r}_{L}) - V_{MM}(\vec{r}_{L}) 
\end{equation}

The main advantage is simplicity --- no explicit MM-ML coupling terms are needed.
The disadvantage is that the interaction between the MM and ML regions is handled entirely at the MM level; the limitations of classical force fields apply to non-bonded interactions.

This approach has been applied successfully and improved binding affinities \cite{Rufa2020-na, Galvelis2023-ks}, yet there is some evidence that for solvation free energies it yields less convincing results \cite{Karwounopoulos2024-fe}.
In one particularly promising demonstration, \citet{Galvelis2023-ks} show that through a combination of well-optimized software and a mechanical embedding ML/MM scheme they are able to perform MD simulations of large solvated protein-ligand complexes ($\sim$30-60 thousand atoms), where the ligand (48-75 atoms) is treated at the level of the MLFF (ANI-2x\cite{smith2020ani}), with computational efficiencies just 1 order of magnitude slower than when solely MM is employed.
It remains to be seen if bespoke fitting of small molecule MM force fields to ML force fields can deliver the same results with increased speed~\cite{horton2022open}. 

An improvement of the coupling (i.e., mutual polarization) of the charge densities between the MM and ML region is called \textit{electrostatic} embedding, in which the interaction of the polarizable ML potential with the rigid MM densities (note that the charges in the MM regions are still fixed) are learned \cite{Lier2022-pt, emle}. 
Nevertheless, improvements in computed free energies are yet to be demonstrated over the simpler mechanical embedding approach.

\paragraph{Mixing time scales.}

One general approach to speeding up MD simulations is to employ multiple time step (MTS) algorithms\cite{Grubmuller1991generalized,Tuckerman1992reversible,Streett1978multiple,Minary2004long,Batcho2001optimized} whereby the slower motions in a system are integrated less frequently than the faster motions. For standard FFs, this timescale separation is typically made with regards to the non-bonded vs. bonded interactions, thereby allowing one to employ a larger time step between the more computationally expensive evaluations of the system's non-bonded interactions. For \textit{ab initio} MD (AIMD) simulations, the same timescale separation cannot be as cleanly made. Instead, AIMD simulations typically adopt an MTS scheme where two levels of treatment for interactions in the system are employed, the target \textit{ab initio} electronic structure method and another more computationally affordable model for the system's potential energy surface, such that the difference in forces between the two levels is slowly varying and can be integrated with a larger \textit{outer} time step while the cheaper model is evaluated every \textit{inner} time step~\cite{Guidon2008abinitio,Steele2013multiple,Luehr2014multiple}. In essence, the cheaper model needs to be accurate, with respect to the target \textit{ab initio} method, for the faster short range interactions. Similar MTS schemes can be used to accelerate both MLFF and ML/MM simulations, and recently \citet{Inizan2023scalable} demonstrated that for their benchmark modeling of benzene in water they were able to speed up their ML/MM (ANI-2x/AMOEBA) simulations approximately 8-fold by employing MTS, with ML/MM and AMOEBA evaluations performed every 2 and 0.5~fs, respectively. 

\subsection{Ecosystems of molecular dynamics simulations, unite?}
\label{subsec:ecosystem}

Currently, the MM infrastructure and simulation platform are typically segregated from the tensor-accelerating frameworks that are ubiquitous in all schools of machine learning and scientific computing.
The reason can be attributed to the particularity of the functional forms and, consequently, the highly specialized kernels designed for these functions.
The intensive requirements for domain knowledge might also justify the need for domain-specific software.
As such, optimizing MM force fields requires reimplementation thereof and to blend MM force fields in an MLFF is also highly non-trivial.

\paragraph{Differentiable simulation.}
To control the course of the simulation and use gradient-based methods to optimize the force field parameters (See Section~\ref{sec:desiderata}),  apart from \textit{post-hoc} reweighting-based techniques~\cite{wang2013systematic, boothroyd2022open} that introduces additional error, making the MD simulation differentiable~\cite{wang2020differentiablemolecularsimulationscontrol, greener2021differentiable, wang2023dmff} is the conceptually simplest avenue.

The technical cornerstone to enable this endeavor is the \textit{adjoint sensitivity} method for taking derivatives across ordinary or stochastic differential equations (ODE/SDE)~\cite{kidger2021hey, chen2018neuralode, pontryagin2018mathematical}, which allows the differentiation of the loss function $\mathcal{L}(\mathbf{x}_T)$, dependent upon a later state $\mathbf{x}_T$ in a trajectory $\mathbf{x}_{0 \cdots t \cdots T}$, to be evaluated in constant memory and linear time.
Specifically, in an ODE setting where the dynamics is controlled by a parametrized function $\operatorname{d} \mathbf{x} / \operatorname{d} t = f(\mathbf{x}; t, \theta)$, for example, the \textit{adjoint}, defined as $a(t) = \partial \mathcal{L} / \partial \mathbf{x}_t$, can be calculated by another ODE:
\begin{equation}
\operatorname{d} a(t) / \operatorname{d} t = 
\-a(t)^T \partial f / \partial \mathbf{x}.
\end{equation}
The gradient w.r.t. to the parameters can then be calculated as:
\begin{equation}
\operatorname{d} \mathcal{L} / \operatorname{d} \theta
=-\int \operatorname{d} t a^T(t) \partial f / \partial \theta,
\end{equation}
where $a \partial f / \partial \mathbf{x}$ and $a^T \partial f / \partial \theta$ can be efficiently computed as vector-Jacobian products.

\paragraph{MM re-implementation in tensor-accelerating frameworks.}
\citet{jaxmd2020, doerr2021torchmd} strive to overcome this barrier by implementing the MM energy functions and sampling strategies in JAX~\cite{jax2018github} and PyTorch~\cite{paszke2019pytorch} to enable the end-to-end differentiation of MM energy functions.
To efficiently do so is met with a multitude of challenges:
to start with, tensor-accelerating frameworks are designed to run a wide variety of hardware platforms, with different support characteristics and capabilities.
For example, the support for different or mixed precisions differs for each piece of hardware, which is not detrimental to ML applications due to its over-parametrized nature~\cite{Mixed-Precision}, but is crucial for MM simulations to run efficiently.
More fundamentally, tensor-accelerating frameworks focus on general applicability and being able to efficiently evaluate a wide range of energy functional forms, with emphasis especially on highly parallel linear algebra operations, whereas MM platforms have more aggressively optimized the few local and pairwise terms.
Efforts~\cite{timemachine} to push towards accelerating MM implementation in tensor-accelerating frameworks usually end up re-writing all lower-level kernels for energy evaluation.

\paragraph{ML plug-in in MM platforms.}
Conversely, \citet{eastman2023openmm, Galvelis2023-ks} offers functionalities to plug MLFF into traditional MD simulations to streamline the inference and sampling of MLFF on biologically relevant systems.
We envision that this will be the drive of the next generation of hardware-specific (such as Anton~\cite{dror2011anton, shaw2021anton}) revolution.
The speed bottleneck in these implementations usually lies in the constant transfer of coordinates and forces between the MM platform and the tensor-accelerating framework.
At the very least, zero-copy is required to avoid duplicating the data or carrying the data from GPUs to hosts.

\subsection{Foundation models for force fields and more.}
Foundation models are an approach to representation learning that involves training a large artificial neural network on extremely large amounts of heterogeneous, multi-modal, and easily available data~\cite{bommasani2022opportunitiesrisksfoundationmodels}. 
The learned representation is then leveraged and fine-tuned for various downstream tasks. 
Training on large and diverse datasets allows the model to generalize well across tasks and domains. Foundation models have enabled truly spectacular achievements in natural language processing and computer vision, and have begun to penetrate the physical sciences as demonstrated by recent advances in protein structure prediction.

Recent work in the field of ML atomistic models has begun to recognize the need for more scalable approaches resulting in the generation of larger datasets, e.g. the Open Catalyst (OC20)~\cite{Chanussot_2021} and Materials Project~\cite{jain2013commentary} datasets, attempts to establish neural scaling laws for chemical models (on single data sources and, typically, on rather small scales) as well as universal models capable of few-shot, or even zero-shot, learning. 
Although far from the successes of natural language processing and computer vision, these results suggest that by scaling these domain-specific models we should be able to achieve comparable successes to those of more traditional computer science domains.

The high dimensionality of the chemical space and the scarcity and computational cost of generating training data, as well as the inconsistencies among available data, represent major obstacles for developing foundation models that are predictive over a broad range of materials and properties as needed for molecular and materials discovery.
To fulfill this vision, critical limitations need to be overcome, such as \textit{data scarcity} and \textit{data redundancy}, the design of novel training strategies that can deal with inconsistencies across datasets, the exploration of adaptive batching and curriculum learning strategies, the development of scalable physics-informed models, as well as the characterization of scaling laws for domain-specific neural architectures.

\section{Conclusions}
Currently, the MM force fields are arguably fast enough but not sufficiently accurate; the ML force fields are accurate enough but painfully slow.
We anticipate that a force field faster than current state-of-the-art MLFF and significantly more accurate than MM, will show greatest utility in the \textit{in silico} modeling of biomolecular systems.
With that in mind, we have herein surveyed the recent advances in either school in search of opportunities to bridge the design spaces.

We envision that the next generation of MLFFs with the desired balance between speed and accuracy will be composed of simple operations such as dot products while able to universally approximate all $E(3)$-invariant functions.
Physically inspired inductive biases will be encoded in this model stability and smoothness is guaranteed.
This model will likely still rely on automatic differentiation and will be implemented in general tensor-accelerating frameworks.
A community-wide effort to design and generate high-quality datasets will ensure its wide chemical space coverage.

\section{A nihilist epilogue: but do we really need a force field?}
Echoing the opening paragraph of this review, computational chemists and biophysicists, on most occasions, are interested in \textit{sampling from} the Boltzmann distribution (Equation~\ref{eq:boltzmann}), rather than just knowing its exact value up to a normalizing constant.
It has quickly emerged to become a focus of cutting-edge research to generate samples directly on the coordinate space for biomolecular systems of interest---from small molecule conformer generation~\cite{Zang_2020, jin2019junctiontreevariationalautoencoder, shi2020graphafflowbasedautoregressivemodel, Mansimov_2019, hoogeboom2022equivariantdiffusionmoleculegeneration}, folded protein structure prediction~\cite{jumper2021highly, abramson2024accurate, ahdritz2024openfold}, to trajectory forecasting~\cite{wang2023spatial, klein2023timewarptransferableaccelerationmolecular}.
More fascinatingly, a class of machine learning models known as \textit{Boltzmann generators}~\cite{noe2019boltzmann, tuckerman2019machine, klein2024transferableboltzmanngenerators, kohler2020equivariant, midgley2024se3equivariantaugmentedcoupling, satorras2022enequivariantnormalizingflows} sample the Boltzmann distribution in an asymptotically unbiased fashion in one shot, without relying on simulation.

\textit{Sampling} and \textit{force field} have long been two orthogonal axes in computational chemistry research, with little cross-disciplinary communications.
The theoretical and experimental advances in energy-based models~\cite{lecun2006tutorial, song2021train} (EBM) and diffusion models~\cite{croitoru2023diffusion, yang2023diffusion, song2021scorebasedgenerativemodelingstochastic, ho2020denoisingdiffusionprobabilisticmodels} should remind ourselves that these can be the same thing.
The loss function in probabilistic generative modeling is usually some rendering of the energy function---in the context of learning by examples, the loss function can be viewed as the local Gaussian or Laplacian extrapolation around the learning set.
When training a generative model, one might get a force field for free~\cite{arts2023onediffusionmodelsforce}.




\section*{Disclosures. }
YW has limited financial interests in Flagship Pioneering, Inc. and its subsidiaries.

\section*{Acknowledgements. }
This research was carried out on high-performance computing resources at Memorial Sloan Kettering Cancer Center and the Washington Square and Abu Dhabi campuses of New York University.

\section*{Funding. }
YW acknowledges support from the Schmidt Science Fellowship, in partnership with the Rhodes Trust, and the Simons Center for Computational Physical Chemistry at New York University. S.M. acknowledges NSF Award 2311632 and the Simons Center for Computational Physical Chemistry for financial support.

\section*{References}
\bibliography{main}

\end{document}